\documentstyle[aps,preprint,tighten,epsf]{revtex}

\def\secder#1#2{{\partial^2 #1\over\partial^2 #2 }}
\begin{document}

  \draft

  \title{A numerical study of the RG equation for the deformed $O(3)$
  nonlinear sigma model}

  \author{L. Belardinelli and C. Destri\footnote[2]{E-mail address:
  destri@milano.infn.it}}
  \address{Dipartimento di Fisica, Universit\`a di Milano and INFN,
  Sezione di Milano, Italy}
  \author{and}
  \author{E. Onofri\footnote[3]{E-mail address:
  onofri@parma.infn.it}}
  \address{Dipartimento di Fisica, Universit\`a di Parma and INFN,
  Gruppo Collegato di Parma, Italy}

  \maketitle

  \begin{abstract}
  The Renormalization Group equation describing the evolution of the
  metric of the nonlinear sigma model poses some nice mathematical
  problems involving functional analysis, differential geometry and
  numerical analysis. In this article we briefly report some results
  obtained from the numerical study of the solutions in the case of a
  two dimensional target space (deformation of the $O(3)$
  $\sigma$-model). In particular, our analysis shows that the so-called
  {\em sausages} define an attracting manifold in the $U(1)$-symmetric
  case, at one loop level. Moreover, data from two--loop evolution are
  used to test the association \cite{FZO} between the so--called
  $SSM_{\nu}$ field theory and a certain $U(1)$-symmetric, factorized
  scattering theory (FST).

  \end{abstract}

  \narrowtext

  \section{Renormalization Group equation}

  The perturbative renormalization of the non linear $\sigma$-model \cite{Fri}
  gives rise to a deformation of the metric according to the (one--loop)
equation
  \begin{equation}
  \frac{{\rm d}g_{ij}}{{\rm d}t}\:=\:-\frac{1}{4\pi}R_{ij}+O(R^2)\; .
  \end{equation}
  This second order non linear partial differential equation (PDE) has been
studied
  in the simplest case of 2-dimensional target manifolds in ref. \cite{FZO}. A
  whole family of solutions is known for the topology of the sphere $S^2$ or
for
  the torus. In this letter we consider the case of $S^2$ only, but our method
  is easily adapted to the toroidal case. As is well known we can introduce
  local coordinates $\{y,\varphi\}$ in such a way that the metric is
conformally
  flat:
  \begin{equation}
  g_{ij}\:=\:e^{\phi}\delta_{ij}
  \end{equation}
  and the RG equation valid to all loop reduces to a single non linear PDE,
  \begin{equation}
  \frac{\partial\phi}{\partial t}\:=\:\beta(R)
  \end{equation}
  where $\beta(R)$ is the exact $\beta-$function and
  \begin{equation}
  R\:=\:-e^{-\phi}\left[\secder\phi y+\secder\phi\varphi\right]
  \end{equation}
  is the scalar curvature.

  A family of solutions in the case of one--loop, namely $\beta(x)=-x/4\pi$ has
been
  presented in \cite{FZO}. These can be obtained starting from the {\em ansatz}
  \begin{equation}
  \phi(y,\varphi;t)\:=\:-\log[a(t)+b(t)f(y)+c(t)h(\varphi)]\; .
  \end{equation}
  The only solutions of this kind which can be extended to $t\rightarrow
  -\infty$ (UV limit) without encountering singularities and which
  exibits a residual $U(1)$ symmetry are the so-called {\em sausage}
  solutions, parametrized by a single real constant $\nu$:
  \begin{eqnarray} \label{salsfam}
  \phi(y,t)&=&-\log(a(t)+b(t)\cosh2y)  \\ \nonumber
  a(t)&=&\frac{1}{2}\nu\coth\left[\frac\nu{2\pi}(t_0-t)\right]\; , \quad
  b(t)=\frac{\nu}{2\sinh\left[\frac\nu{2\pi}(t_0-t)\right]} \; .
  \end{eqnarray}
  These trajectories in the space of all metrics are believed to be the
one--loop
  approximation of some integrable deformation of the costant--curvature
  trajectory ($O(3)$ $\sigma$-model), to which they tend as $\nu\rightarrow 0$.
  To discuss the general solution of the RG equation we have to construct  a
  numerical integration algorithm. The first problem is posed by the
  divergence of the conformal factor as $|y|\rightarrow\infty$: $\phi\sim
-2|y|$
  as $|y|\rightarrow\infty$. This general property (which follows directly from
  Gauss--Bonnet theorem) causes the factor $\exp(-\phi)$ to diverge, hence
  amplifying the numerical error of the differential term at large $|y|$. To
  overcome this difficulty we introduce a background field $\phi_0(y,t)$, and
we
  consider the equation for the shifted field $\phi=\eta+\phi_0$. The
background
  field is conveniently chosen as the constant curvature solution
  \begin{equation}
  \phi_0\:=\:\log\left[{A(t)}\over{4\pi\cosh^2y}\right] \; ,
  \end{equation}
  where according to the RG equation $A(t)$ satisfies
  \begin{equation}
  \dot{A}(t)A(t)^{-1}\:=\:\beta(8\pi A(t)^{-1})\; .
  \end{equation}
  Introducing a new evolution parameter (``time'') $\tau$, defined by
  \begin{equation}
  A(t)\:=\:4\pi e^{-\tau}
  \end{equation}
  the RG equation expressed in terms of the shifted field $\eta$ reads
  \begin{equation}        \label{eq:RGeq}
  \partial_{\tau}\eta\:=\:1-\frac{\beta(\tilde{R}(\eta))}{\beta(2e^{\tau})} \\
  \end{equation}
  with
  \begin{equation}
  \tilde{R}(\eta)\equiv
  R(\phi_0+\eta)\:=\:e^{\tau-\eta}(2-\Delta_0\eta)\; .
  \end{equation}
  Here $\Delta_0\equiv\cosh^2\!y\,\partial^2_y+\partial^2_{\varphi}$ is the
  standard $O(3)-$invariant Laplacean on the sphere. The original time
  scale can be simply recovered by quadratures
  \begin{equation}   \label{quadr}
  t_0-t\:=\:\int^{\tau}_0\frac{{\rm d}\tau'}{\beta(2e^{\tau'})}\; .
  \end{equation}
  At this stage we restrict our attention to $U(1)-$symmetric solutions
  $\eta=\eta(y,t)$.

  \section{Numerical integration of the RG equation}

  At this point it is convenient to adopt the standard spherical
  coordinates by letting $y=\log(\cot(\vartheta/2))$.  To get a good
  accuracy in the evaluation of the Laplacean we apply the spectral
  method, that is we expand $\eta$ in Legendre polynomials
  $P_n(\cos\vartheta)$, which are the eigenfunctions of $\Delta_0$.
  Thus, we developed the following finite--dimensional implementation of
  the standard expansion in Legendre polynomials.

  Let $x_j^{(L)}$ be the zeroes of the $L$-th Legendre polynomial
  $P_L(x)$; we adopt $\{x_j^{(L)}\:|\:j=1,\dots,L\}$ as our finite
  grid\footnote {Solutions with reflection simmetry $(x\rightarrow -x)$,
  such as the sausage solutions, can be studied by a restriction to
  even--order polynomials; our algorithm can be easily adapted to this
  case, with a gain of a factor four in memory requirements} and sample
  the field $\eta(y,t)$ at the image points $y^{(L)}_j$. \\ The finite
  Legendre expansion is then realized as follows:
  \begin{eqnarray}
  &&\eta(x)\:=\:\sum_{l=0}^{\infty}\eta_lP_l(x)\approx
  \sum_{l=0}^{L-1}\eta_lP_l(x) \\
  &&\eta_l=\frac12\int^1_{-1}{\rm d}x\;\eta(x)P_l(x)(2l+1)
  \approx\frac12\sum_{j=1}^L\eta(x_j^{(L)})P_l(x_j^{(L)})
  w_j^{(L)}(2l+1) \equiv(\cal L\eta)_l
  \end{eqnarray}
  where $w_j^{(L)}$ are the Gaussian integration weights \cite{Hoc} for
  Legendre polynomials.  This finite expansion allows us to represent
  the Laplacean {\em exactly} on polynomials of degree less than $L$,
  since the coefficients $\eta_l$ are exact in this case. The lack
  of a fast implementation analogous to FFT limits our algorithm
  in practice to $L\sim 200$ on current workstations, but this proves to
  be adequate for our purposes.

  Having determined the finite transform $\cal L$ on the basis of
  Legendre polynomials, the action of the Laplacean $\Delta_0$ is
  represented by a matrix $\cal L^{-1}\Lambda\cal L$, where $\Lambda$ is
  diagonal, with eigenvalues $\{-l(l+1)\:|\:l=0,\dots,L-1\}$. In terms
  of this representation is quite easy to compute the spectrum of
  zero--modes of the field $\phi$, a problem considered in
  ref. \cite{FZO}. We just have to diagonalize the finite matrix
  $-(1/2)\Delta+(1/8)\tilde{R}$, where
  $\Delta=\exp(\tau-\eta)\Delta_0$. The results agree with the
  previously computed ones for the sausages \cite{BT} (notice however
  that the present method is much simpler and the spectrum can be
  computed in parallel with the RG evolution).

  With the finite Legendre--transform algorithm at hand, we can now
  consider the integration of eq.~(\ref{eq:RGeq}). We can work to any
  loop, provided we know the corresponding $\beta-$function. We have
  implemented the algorithm in {\sc matlab} \cite{BO} which provides
  efficient routines of dia~-go~-nalization and of adaptive-step
  integration. The accuracy of the code has been tested on the known
  one--loop sausage solution, attaining a typical maximal deviation of
  $1$ part in $10^9$ over a time interval $-3<\tau<2$ and $\nu<.25$. In
  Fig.1 we show a typical one--loop sausage evolution; to aid the
  visualization we have recovered the embedding of the surface in $R^3$
  in such a way that the induced metric coincides with
  $\exp(\phi)\delta_{ij}$. The accuracy is limited essentially from the
  large eigenvalues of the Laplacean which grow as the square of the
  finite grid dimension. Moreover as $\nu$ and/or $-\tau$ increase the
  curvature tends to be confined at the extremities of the sausage,
  which requires finer and finer grid. Presently, up to $200$ points, we
  cannot go beyond $\tau\approx -4$ for $\nu\approx .25$, but there is
  no limit in principle.

  The algorithm can now be applied to investigate, at one--loop, the
  existence of {\em attracting manifolds}, in the space of all
  metrics. A conjecture of Fateev and Zamolodchikov states that the
  sausages constitute a stable manifold and all other geometries
  converge to some sausage, parametrized by the real $\nu$, in the IR
  direction $\tau\rightarrow\infty$. \\ This fact manifests itself
  quite clearly in our numerical data. We may define a distance
  function by \begin{equation}
  \mathrm{dist}(\eta_1,\eta_2)\:=\:\int{\rm d}x[\tilde{R}(\eta_1)-
  \tilde{R}(\eta_2)]^2 \end{equation} and measure the distance to
  the sausage hypersurface by \begin{equation}
  D(\tau)\:=\:\inf_{\nu}{\mathrm{dist}}(\eta(\tau),\eta_{\nu}(\tau))
  \end{equation} where $\eta_{\nu}$ is the sausage solution specified
  by $\nu$. The data show a clear exponential decay $D(\tau)\approx
  A\exp(-m\tau)$, with $m\approx 8$ regardless of the initial surface
  (see Tab.1). For any given starting geometry we record the value
  $\nu_{\mathrm{eff}}$ where the infimum is reached; its limit
  $\nu_{\mathrm{lim}}$ as $\tau\rightarrow\infty$ gives a definition
  of {\em sausageness} of any given surface. For instance, given an
  ellipsoid with cylindrical symmetry and eccentricity $\epsilon$ we
  can measure $\epsilon(\nu_{\mathrm{lim}})$, at one--loop order (see
  Fig.2). This shows a remarkable property: $\epsilon$ is a universal
  function of $s=\nu_{\mathrm{lim}}\frac{e^{-\bar{\tau}}}{2}$, where
  $\bar{\tau}$ is the starting scale of the evolution. This function
  seems to be $\epsilon=\tanh s$, at least for small $s$. \\

  \section{Variational equations and stability}

  Another way to discuss the attracting nature of the sausage
  manifold is to study the Jacobi variational equations
  around the sausage solution.  The linearized equations take on the
  following form ( calling $\chi$ the variation): \begin{equation}
  \partial_{\tau}\chi\:=\:H(\eta,\tau)\chi \end{equation} where
  \begin{equation}
  H(\eta,\tau)\:=\:\frac{\beta'(\tilde{R}(\eta))}{\beta(2e^{\tau})}
  \left(e^{\tau-\eta}\Delta_0+\tilde{R}(\eta)\right)\; .
  \end{equation}
  The spectrum of $H$ is not {\sl a priori} of much
  significance, since the evolution equations are
  time--dependent. However, if we rely on the adiabatic approximation,
  the spectrum is directly related to stability. Applying the
  finite Legendre--transform $\cal L$ the spectrum of $H$ is easily reduced to
  that of a Hermitean matrix; choosing for $\eta$ some sausage
  $\eta_{\nu}$ we find a single positive eigenvalue denoting an
  unstable mode. As we presently show, this is easily interpreted. \\
  It is immediate to find that the Jacobi variational equation at one
  loop:
  \begin{equation}
  \partial_{\tau}\chi\:=\:\frac{1}{2}\left(\exp(-\eta_\nu)\Delta_0+e^{-\tau}
  \tilde{R}(\eta_\nu)\right)\chi
  \end{equation}
  admits the two solutions
  \begin{eqnarray} &&\chi_1\:=\:\tilde{R}(\eta_\nu) \\
  \label{varsals} &&\chi_2\:=\:1-(1/2)\exp(-\tau)\tilde{R}(\eta_\nu)\;.
  \end{eqnarray}
  Their existence is not surprising: the family of
  sausage solutions (\ref{salsfam}) depends on two parameters $t_0$ and
  $\nu$, so that the derivatives with respect to $t_0$ and $\nu$ give
  rise to two independent solutions of the variational equations.  Now
  $\chi_1$, which comes from the time derivative represents the
  unstable solution; this kind of instability is of no concern, since
  it corresponds to a simple redefinition of the initial time
  parameter $t_0$ and it can be fixed by restricting to a given
  initial area. The other solution $\chi_2$ comes from the
  $\nu-$derivative and is tangent to the manifold of sausages. The
  component of a generic variation $\chi$ which is orthogonal to both
  $\chi_1$ and $\chi_2$ provides a measure of the distance of a
  generic solution from the sausage manifold. \\

  Numerical results (Tab.2 and 3) show that the orthogonal part is
  exponentially decreasing with a slope ($\approx 8.73$)\footnote{This
  slope is somewhat higher than the one found in sect. II; there we
  were actually considering variations with a nonzero slowly--decaying
  component along $\chi_2$.} much higher than the slope of $\chi_2$,
  which from equation (\ref{varsals}) is found to be 2. The point is that
  there is an overall convergence in the infrared toward the constant
  curvature metric. All sausages converge to a sphere with vanishing
  radius, forcing the decay of the mode $\chi_2$; however, the rate of this
  convergence is slower than the rate of decay of all other
  modes. Fig.3 shows how the longitudinal component (plotted in the
  vertical direction) is stil large when the transversal one is already
  negligible. In Fig.4 there is an impressive, though only qualitative
  proof of what we found (one should compare Fig.4 with Fig.1, where the
  sausage evolution is plotted).

  Conversely, the evolution in the ultraviolet direction $\tau\rightarrow
  -\infty$ is strongly unstable, which makes it quite hard to follow
  numerically; all truncation errors are chaotically amplified and the
  calculation becomes rapidly unreliable.

  \section{The free energy of the sausage sigma model}

  In ref. \cite{FZO} the model corresponding to $U(1)-$symmetric,
  one--loop deformation of the $O(3)$ sigma model (termed $SSM_{\nu}$
  model) is associated with a $U(1)-$symmetric deformation of the $O(3)$
  FST. To test the correctness of the identification, some physical
  quantities calculated from $S$-matrix data via Thermodynamic Bethe
  Ansatz tecniques are compared in the UV region with the corresponding
  quantities obtained perturbatively from the one--loop sausage action
  \begin{equation}     \label{eq:sausaction}
  \cal A_{SSM_{\nu}}\;=\;\int
 d^2x\frac{(\partial_{\mu}y)^2\:+\:(\partial_{\mu}\varphi)^2}{a\:+\:b\cosh2y}\;
{}.
  \end{equation}
  Their non--trivial matching is a strong (though indirect)
  proof of the correctness of the association.

  However, while for the FST we have exact (at least in principle)
  results, for the field theory we are strongly limited by the fact that
  the ``sausage metric'' is a solution to the RG equation {\em only to
  one--loop}. So, it would be very important to have higher--loop
  solutions of the RG equation with the same features of the sausage, so
  to confirm (or deny) the validity of the association proposed in
  \cite{FZO}.  This is not to be taken for granted. To any relativistic
  FST one would like to associate an integrable QFT; however, even if
  one assumes that the action (\ref{eq:sausaction}) is the one--loop
  approximation of an exact trajectory of the RG group, which would
  define the {\em exact} sausage model, there is no {\em a prori} reason
  to expect such model to be integrable. In particular, $S^2$ with any
  $U(1)$-symmetric metric {\em is not} a symmetric space and one cannot
  use the results of refs. \cite{Eich}\cite{Pol} to establish the
  classical and/or quantum integrability of the model.

  Considering the zero--temperature free energy $f(A)$ in the presence
  of a constant external field $A$, the one--loop calculation based on the
  action (\ref{eq:sausaction}) gives the result
  \cite{FZO}
  \begin{equation}  \label{eq:freeenergy}
  f(A)=-A^2\:e^{\phi(y=0,t)}\; ,
  \end{equation}
  where $t_0-t=\log A/\Lambda$, $\Lambda$ is a subtraction scale and one
  works in the ``scaling limit'':
  \begin{equation}
  \nu\rightarrow 0 \;,\quad t \rightarrow -\infty \;,\quad\nu t\;\;
\mathrm{fixed}
  \end{equation}
  This limit serves to eliminate all higher--loops contributions, which
  cannot be properly taken into account using the one--loop action
  (\ref{eq:sausaction}).  To obtain higher--loop correction to
  (\ref{eq:freeenergy}) one needs the corresponding higher--loop
  solution of the RG equation. Formula (\ref{eq:freeenergy}) however remains
  valid, assuming as natural that the higher--loop action is an even,
  convex function of $A$.

  With our algorithm, the quantity in (\ref{eq:freeenergy}) is evaluated as
  \begin{equation}  \label{freenum}
  f(A)_{\mathrm{num}}=-A^2e^{\eta(t,x_{\mathrm{med}})}e^{\phi_0(t,y=0)}
  =-A^2e^{\eta(t,x_{\mathrm{med}})}\frac{t_0-t}{4\pi}
  \end{equation}
  where $x_{\mathrm{med}}$ is the middle point of the finite grid with
represents
  the coordinate $x=\cos\theta$.  The two--loop $\beta-$function is well
  known \cite{Fri}: $\beta(x)=-x/4\pi-x^2/(4\pi)^2$. Lacking an exact,
  $U(1)-$symmetric two--loop solution of the RG equation, the first
  thing one could try is a linerization around the one--loop sausage
  solution, regarding $\nu$ as perturbative parameter. Notice in fact that
$\nu$
  may be scaled out from the sausage solution, through the substitutions
  \begin{equation} \label{scalnu}
  \phi\rightarrow\phi+\log\frac{\nu}{4\pi}\;,\quad t\rightarrow \nu t
  \end{equation}
  which leave the one--loop equation invariant (actually, with our
  choice, the scalar curvature $R$ is conveniently scaled by a factor of
  $4\pi$). On the other hand the two--loop equation is transformed by
  (\ref{scalnu}) to
  \begin{equation}
  \frac{\partial\phi}{\partial t}\:=\:-R-\nu R^2
  \end{equation}
  where the last term could now be treated perturbatively for small
  sausage deformations. However, even the linearized equation proves to
  be too difficult to solve exactly, when the correct boundary
  conditions for $\phi$ are taken into account, and we resort to a
  numerical approach.

  Our idea is the following: in the scaling limit (far UV
  and small deformations), the exact solution must have the same
  features of its various loop--wise approximations.  We can now study
  with our algorithm the two--loop evolution, using as initial data
  in the far UV (say at $t=\bar{t}$)
  a metric with the ``sausage features '' (e.g. a sausage itself,
  one--loop solution).  We then evaluate our physical quantity in a
  fixed scale interval $(t_1,t_2)$ and we compare it with the same
  quantity evaluated starting from the FST by means of Bethe Ansatz
  techniques \cite{FZO}\cite{HMN}. {\em We expect that the difference
  between the two quantities tends to zero when the starting scale of
  the evolution is pushed farther and farther in the UV}. At one--loop
  level this is easily verified, using as starting data, e.g., the
  function
  \begin{equation}
  \phi(\bar{t},y)=\phi_0(\bar{t},y)+\log\frac{2}{\nu}-
  \log(1+2e^{\nu(\bar{t}-t_0)/2\pi}\cosh
  2y+ e^{\nu(\bar{t}-t_0)/\pi})
  \end{equation}
  The above mentioned difference vanish exponentially when the initial
  scale $\bar{t}$ is sent to $-\infty$.

  In ref. \cite{FZO} the Bethe Ansatz techniques are used to compute the
  free energy as a function of the chemical potential $h$ coupled to the
  physical particles. The input is the factorized, $U(1)-$symmetric
  $S$-matrix with deformation parameter $\lambda$ relative to the
  $S$-matrix of the $O(3)$ sigma model. The output reads, in the scaling
  limit $\lambda\to 0$, $h\to\infty$, $\lambda\log h$ fixed,
  \begin{equation}  \label{eq:nonp2loopfren}
  f(h)=-\frac{h^2}{2\pi\lambda}\frac{1-u}{1+u}\left[1+4\lambda\frac
  {u}{1-u^2}\log\frac{1-u}{2\lambda}+O(\lambda^2\log^2\lambda)\right]
  \end{equation}
  where $u=\left(\frac{me^{3/2}}{8h}\right)^{2\lambda/(1-\lambda)}$, and
  $m$ is the mass of the physical particles.
  The connection among the various quantities in~(\ref{eq:nonp2loopfren})
  and those contained in the perturbative expressions is established as
  follows:
  \begin{eqnarray}
  && \frac{me^{3/2}}{8}=\Lambda+O(\nu)  \label{link1} \\
  && h=A+O(\nu) \label{link2} \\
  && \frac{\lambda}{\nu}=\frac1{4\pi}[1+O(\nu)] \label{link3}
  \end{eqnarray}
  The relation between the ``scale variables'' $z=\nu(t_0-t)/2\pi$ and
  $u$ follows by noticing that in the perturbative calculations we put
  $t_0-t=\log(A/\Lambda)$. Substituting (\ref{link1})-(\ref{link3}) in
  (\ref{eq:nonp2loopfren}) we obtain
  \begin{equation} \label{eq:frentwolooppert}
  \frac{f(A)_{\mathrm{an}}}{A^2}=
  -\frac1{\pi}\left[\frac{2\pi}{\nu}\tanh(z/2)+\frac{\log\nu}
  {\cosh^2(z/2)}-\frac{\log(2\pi(1-e^{-z}))}{\cosh^2(z/2)}+
  O(\nu)\right]
  \end{equation}
  where the subscript ``an'' stands for ``analytical'' (compare
  (\ref{freenum})).  The first term comes from the one--loop action
  (\ref{eq:sausaction}), while the second and the third represent the
  two--loop contributions. The presence of the non--analytic behaviour
  $\nu\log\nu$ ensures the correct $O(3)-$symmetric limit ($\nu\to 0$ at
  fixed $t_0-t$).  Corrections of order $\nu$ are not known; to
  establish them we should know the corrections of the same order to
  eqs. (\ref{link1})-(\ref{link3}). Our aim is now to compare
  expressions $f(A)_{\mathrm{an}}$ and $f(A)_{\mathrm{num}}$.

  We have collected numerical data relative to the two--loop evolution
  of sausages with $\nu$ fixed, starting at different initial scales
  $\bar{\tau}$ (recall that $\tau$ is related to $t_0-t$ by
  (\ref{quadr})).  Fig. 5 reports the results; here $\nu=0.1$ and the
  $\tau$ interval is chosen as $(\tau_1=-3.85,\tau_2=-3.8)$. The
  quantity $\Delta$ is defined as \begin{equation}
  \Delta(\bar{\tau})=\max_{\tau_1<\tau<\tau_2}
  \,|f(A)_{\mathrm{an}}/A^2-f(A)_{\mathrm{num}}/A^2| \end{equation}
  and it is plotted for different values of the initial scale
  $\bar{\tau}$.\\ $\Delta$ {\em does not} decrease when the scale is
  pushed towards far UV; instead, initially increase and then reach an
  asymptotical value.  The value to which $\Delta$ tends is of order
  $10^{-2}$; that means a discrepancy of about 5 parts in 10000
  relative to the value of the free energy. This is well above the
  numerical errors of our algorithm.  \\ We controlled if this
  discrepancy can be reabsorbed in a ``renormalization'' of $\nu$;
  leaving it as a free parameter in (\ref{eq:frentwolooppert}), we
  seeked for a minimum of $\Delta$ in a neighborhood of $0.1$. The
  result was negative. \\ In principle, the non--zero asymptotic value
  of $\Delta$ may be eliminated by the $O(\nu)$ correction; to test
  this hypotesis we repeated the evaluation of $\Delta$ for different
  $\nu$ values, and correspondingly redefined the $\tau-$interval so
  to leave the interval $(\nu t_1,\nu t_2)$ fixed, according to the
  scaling limit. The results are summarized in Fig. 6: the
  asymptotic value of $\Delta$ seems not to depend appreciably upon
  $\nu$. It remains always of the same order of magnitude, and {\em
  does not grow with} $\nu$.

  \section{Conclusions}

  The RG equation for the non linear sigma model with fields taking
  values on a target space with the topology of the 2-sphere is a
  nonlinear partial--differential equation, possibly nonlocal (if the
  $\beta$ function were known to all loop). These features makes it hard
  to caracterize the general solution in an analytical way.  The
  numerical approach appears at the moment to be the only way to obtain
  quantitative informations about the higher--loop solutions. \\
  The spectral method presented in this letter proved to be a useful tool
  for the construction of an efficent algorithm; this remains true even
  if the exact RG equation should turn out to be nonlocal.  Another
  future developement concerns the study of solutions without any residual
  symmetry, hopefully with the help of some ``fast'' algorithm analogous
  to the well known FFT for the Fourier case. \\

  In the one--loop
  approximation we have shown the attractive nature of the
  $U(1)-$symmetric family of solutions, the so--called sausages.

  As far as the two--loop evolution is concerned, however, our data show a
  discrepancy between calculations based on the RG equation and
  analogous calculations based on the $U(1)-$symmetric FST proposed in
\cite{FZO}. \\ This probably means that the identification put forward
  in \cite{FZO} is not correct, in the sense that the $U(1)-$symmetric FST
  after all does not correspond to a $U(1)-$symmetric nonlinear sigma model
  reducing to the sausage at one--loop. On this point, the following
  alternative conjecture seems natural.

  The factorized $S-$matrix put forward in \cite{FZO} is a
  quantum--group deformation of the $O(3)$ $S-$matrix, in the sense
  of, e.g., ref. \cite{Jim}. In practice, the triplet of massive
  particles of the model forms a spin-one irreducible representation
  of $SU(2)_q$ and scatter through a $SU(2)_q-$invariant $S-$matrix
  with $q\simeq e^{i\lambda}$. Hence, the QFT corresponding to such
  FST should enjoy a nonlocal hidden $SU(2)_q$ invariance, whose
  $U(1)$ subgroup coincides with the manifest $U(1)-$symmetry locally
  implemented. In the usual spirit of the sigma models, in which the
  field--theoretical symmetries follow from geometrical properties of
  the target manifold, we expect this quantum--group invariance to
  follow from the noncommutative geometry of a $q-$deformed target
  manifold $SU(2)_q/U(1)$ (attempts in constructing $q-$deformed sigma
  models can be found in refs. \cite{AV} \cite{FLZ}). In other words,
  we suspect that the FST put forward in \cite{FZO} does not
  correspond, beyond the one--loop approximation, to any conventional
  nonlinear sigma model (that is to a model in which the fields take
  values in an ordinary manifold with commuting geometry). It would be
  interesting to explain why there exist a one--loop conventional
  sigma model, perhaps through a suitable expansion near $q=1$. This
  matter clearly requires further investigations.

  \begin{figure} \begin{center} \epsfverbosetrue \leavevmode
  \epsfxsize=12.cm \epsfysize=10.cm \epsfbox{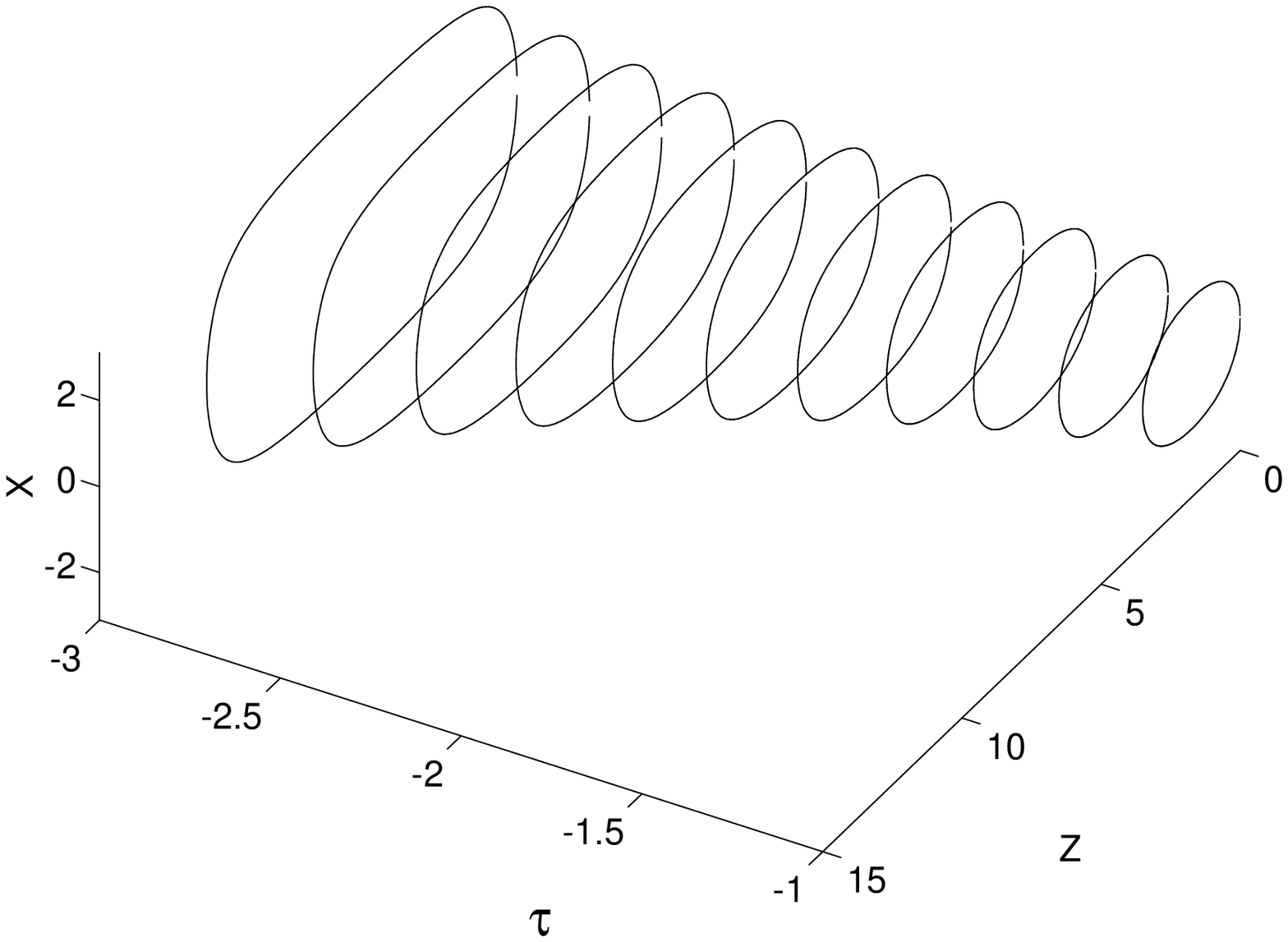}
  \end{center} \caption[One--loop evolution of the sausage solution]
  {{One--loop evolution of the sausage solution}} \label{fig1}
  \end{figure}

  \begin{figure} \begin{center} \epsfverbosetrue \leavevmode
  \epsfxsize=12.cm \epsfysize=10.cm \epsfbox{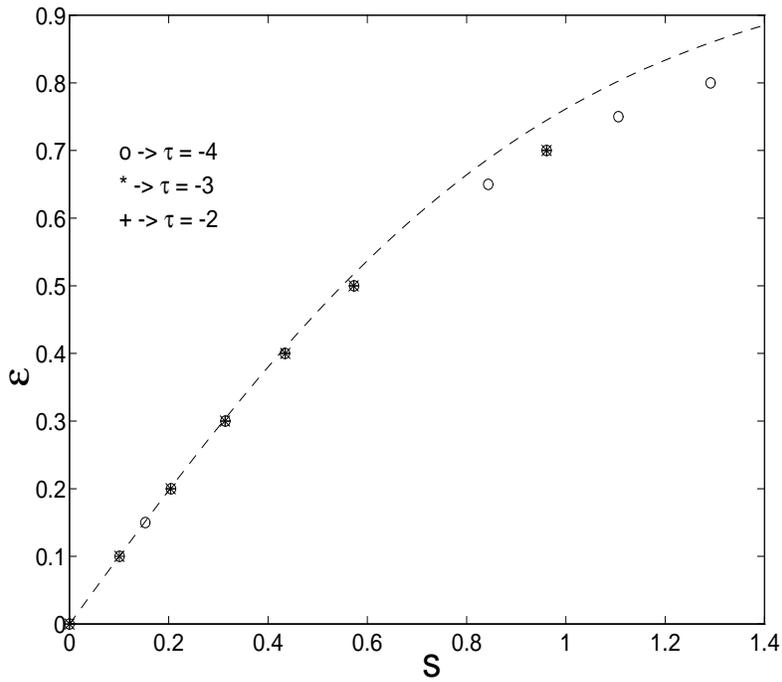} \end{center}
  \caption[The eccentricity as a universal function of
  $s=\nu_{\mathrm{lim}}$. The dashed curve is $\epsilon=\tanh s$]
  {{The eccentricity as a universal function of
  $s=\nu_{\mathrm{lim}}\frac{e^{-\bar{\tau}}}{2}$. The dashed curve is
$\epsilon=\tanh s$}}
  \label{fig2} \end{figure}

  \begin{figure} \begin{center} \epsfverbosetrue \leavevmode
  \epsfxsize=12.cm \epsfysize=10.cm \epsfbox{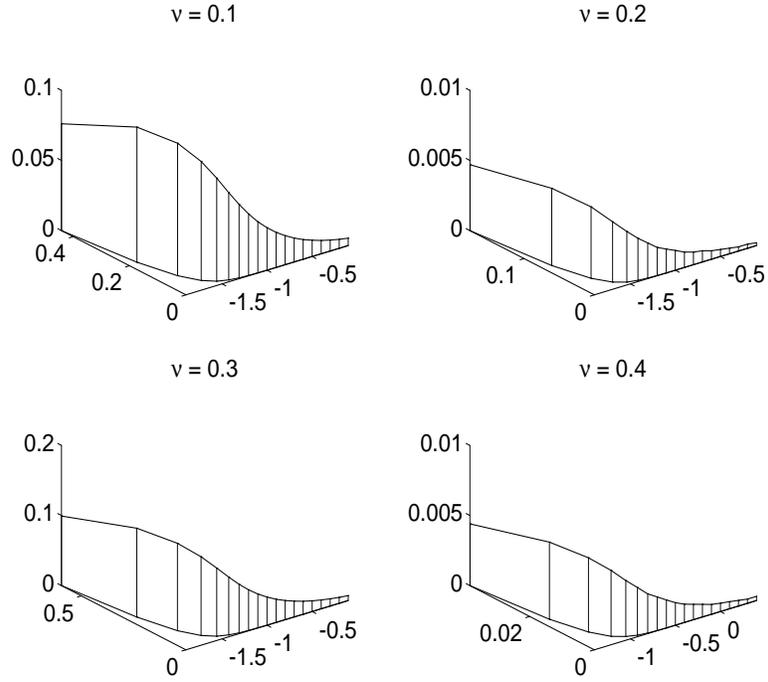} \end{center}
  \caption[Three--dimensional view of the decay of the longitudinal
  and transversal modes] {{Three--dimensional view of the decay of the
  longitudinal and transversal modes}} \label{fig3} \end{figure}

  \begin{figure} \begin{center} \epsfverbosetrue \leavevmode
  \epsfxsize=12.cm \epsfysize=10.cm \epsfbox{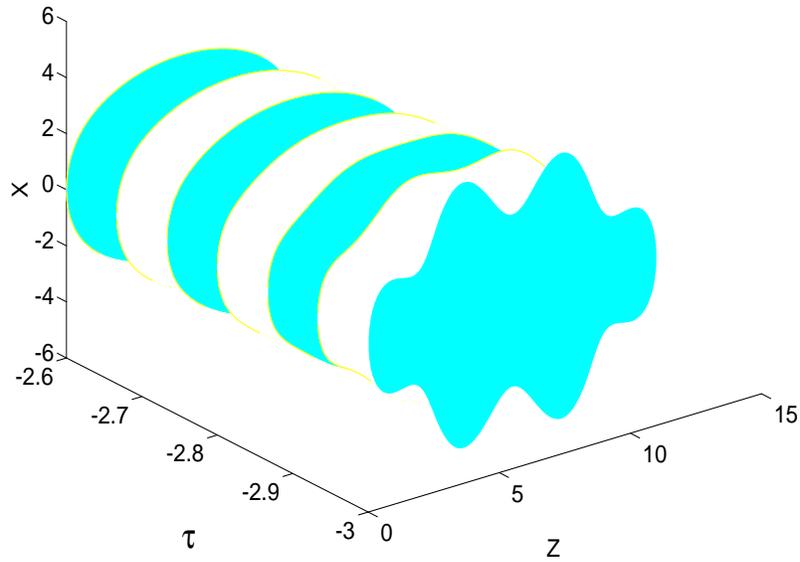} \end{center}
  \caption[A pictorial way of illustrating the attractive nature of
  the sausage solution] {{A pictorial way of illustrating the
  attractive nature of the sausage solution}} \label{fig4}
  \end{figure}

  \begin{figure} \begin{center} \epsfverbosetrue \leavevmode
  \epsfxsize=12.cm \epsfysize=10.cm \epsfbox{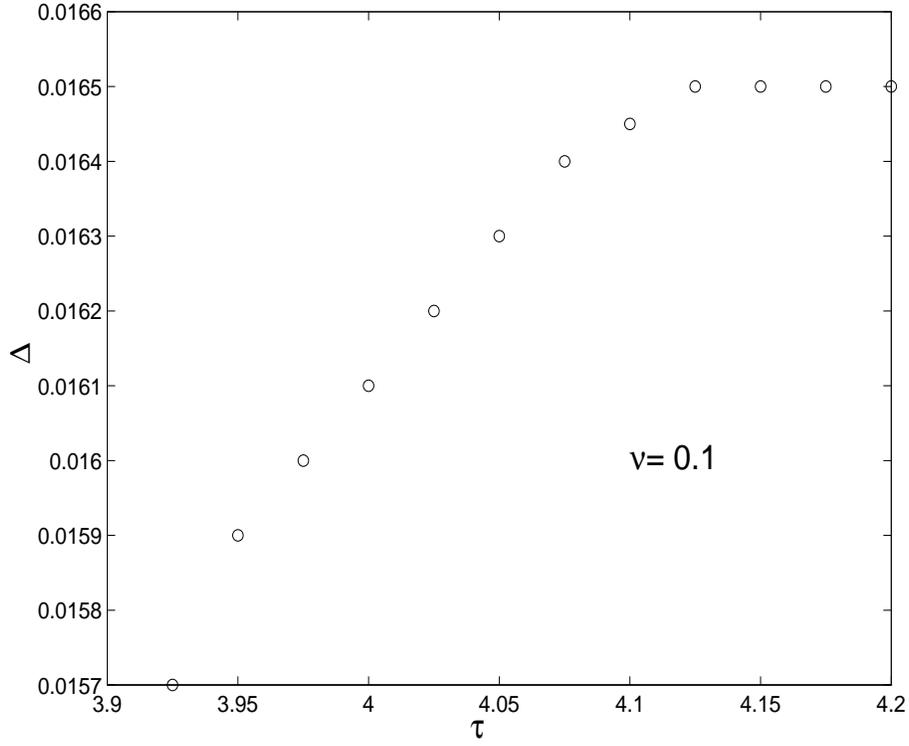} \end{center}
  \caption[The difference, for various values of $\bar{\tau}$, between
  the free energy evaluated from RG equation at two--loop and the
  corresponding quantity evaluated from FST data] {{ The
  difference, for various values of $\bar{\tau}$, between the free energy
  evaluated from RG equation and the corresponding quantity evaluated
  from FST data}} \label{fig5} \end{figure}

  \begin{figure} \begin{center} \epsfverbosetrue \leavevmode
  \epsfxsize=12.cm \epsfysize=10.cm \epsfbox{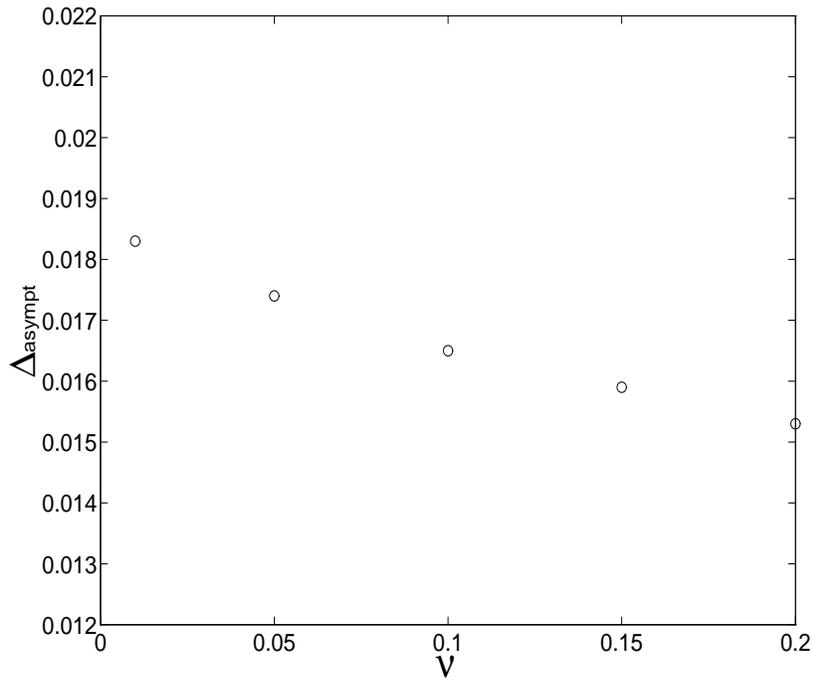} \end{center}
  \caption[The asymptotic value of $\Delta$ for different values of
  the parameter $\nu$] {{The asymptotic value of $\Delta$ for
  different values of the parameter $\nu$}} \label{fig6} \end{figure}

\begin{table}[m]
\caption{Evolution of an ellipsoid; $D$ is the distance to
the sausage manifold.}\label{tab:ellips}
\begin{center}
\begin{tabular}{|c|c|c|} \hline
$\tau$ &  $\nu_{\mathrm{eff}}$ & $D$ \\ \hline
-4.00000 &  0.03761 & 0.04265 \\
-3.90000 & 0.03609 & 0.01453 \\
-3.80000 & 0.03557 & 0.00551 \\
-3.70000 & 0.03536 & 0.00217 \\
-3.60000 & 0.03528 & 0.00088 \\
-3.50000 & 0.03524 & 0.00056 \\
-3.40000 & 0.03523 & 0.00015 \\
-3.30000 & 0.03522 & 0.00006 \\
-3.20000 & 0.03522 & 0.00003 \\
-3.10000 & 0.03522 & 0.00001 \\
-3.00000 & 0.03521 & 0.00000 \\
-2.90000 & 0.03521 & 0.00000 \\
-2.80000 & 0.03521 & 0.00000 \\
-2.70000 & 0.03521 & 0.00000 \\
-2.60000 & 0.03521 & 0.00000 \\
-2.50000 & 0.03521 & 0.00000 \\  \hline
\end{tabular}
\end{center}
\end{table}

%\begin{table}[m]
%\begin{center}
%\caption{The {\it sausageness} of an ellipsoid with eccentricity
%$\epsilon$ and starting time  $\tau=-3$.}\label{tab:ecc}
%\begin{tabular}{|c|cccccc|} \hline
%\small $\epsilon$ & 0.10 & 0.20 & 0.30 & 0.40 & 0.50 & 0.75  \\ \hline
%\small $\nu_{eff}$ & 0.0037 & 0.0075 & 0.0115 & 0.0159 & 0.0210 &  0.0405 \\
%%\hline
%\end{tabular}
%\end{center}
%\end{table}

\begin{table}[m]
\caption{Decay of longitudinal component ($\chi_\parallel$)
and of the transversal one ($\chi_\perp$) at $\nu= .1$}\label{tab:decay1}
\begin{center}
\begin{tabular}{|c|c|c|} \hline
    $\tau$   & $\chi_\parallel $   & $\chi_\perp$ \\ \hline
   -1.9000&    0.0047&    0.1772\\
   -1.8000&    0.0055&    0.0734\\
   -1.7000&    0.0051&    0.0303\\
   -1.6000&    0.0043&    0.0125\\
   -1.5000&    0.0036&    0.0051\\
   -1.4000&    0.0030&    0.0021\\
   -1.3000&    0.0025&    0.0009\\
   -1.2000&    0.0020&    0.0004\\
   -1.1000&    0.0017&    0.0001\\
   -1.0000&    0.0014&    0.0001\\
   -0.9000&    0.0011&    0.0000\\
   -0.8000&    0.0009&    0.0000\\
   -0.7000&    0.0008&    0.0000\\
   -0.6000&    0.0006&    0.0000\\
   -0.5000&    0.0005&    0.0000\\
\hline
\end{tabular}
\end{center}
\end{table}

\begin{table}[m]
\caption{Decay of longitudinal component ($\chi_\parallel$)
and of the transversal one ($\chi_\perp$) at $\nu= .3$}\label{tab:decay3}
\begin{center}
\begin{tabular}{|c|c|c|} \hline
    $\tau$   & $\chi_\parallel $   & $\chi_\perp$ \\ \hline
   -1.9000&    0.0761&    0.4415\\
   -1.8000&    0.0963&    0.2057\\
   -1.7000&    0.0944&    0.0939\\
   -1.6000&    0.0848&    0.0421\\
   -1.5000&    0.0734&    0.0186\\
   -1.4000&    0.0623&    0.0081\\
   -1.3000&    0.0524&    0.0035\\
   -1.2000&    0.0438&    0.0015\\
   -1.1000&    0.0364&    0.0006\\
   -1.0000&    0.0302&    0.0003\\
   -0.9000&    0.0250&    0.0001\\
   -0.8000&    0.0207&    0.0000\\
   -0.7000&    0.0171&    0.0000\\
   -0.6000&    0.0140&    0.0000\\
   -0.5000&    0.0116&    0.0000\\
\hline
\end{tabular}
\end{center}
\end{table}

\end{document}